\documentclass[12pt]{iopart}
\usepackage{amsfonts}
\begin{document}
\title{Ultrarelativistic circular orbits of spinning particles in a
Schwarzschild field}
\author{Roman  Plyatsko}
\address{ Pidstryhach Institute of Applied Problems in Mechanics and
Mathematics\\ Ukrainian National Academy of Sciences, 3-b Naukova
Str.,\\ Lviv, 79060, Ukraine}

\ead{plyatsko@lms.lviv.ua}

\begin{abstract}

 Ultrarelativistic circular orbits of spinning
 particles in a Schwarzschild field described by the
Mathisson-Papapetrou equations are considered. The preliminary
estimates of the possible synchrotron electromagnetic radiation of
highly relativistic protons and electrons on these orbits in the
gravitational field of a black hole are presented

\end{abstract}

\submitto{\CQG}
 \pacs{ 0420, 9530S}

\maketitle
\section {Introduction}

The theory of synchrotron radiation of different types
(electromagnetic, scalar, gravitational) near a Schwarzschild and
Kerr black hole was initiated more than 30 years ago \cite{1,2}.
Two cases of the synchrotron radiation are distinguished: 1) when
a test particle is moving in a highly relativistic geodesic
circular orbit about a black hole, and 2) when an orbit is
circular but nongeodesic, caused by the combined effect of the
gravitational field and other factors (for example, a magnetic
field). In a Schwarzschild field the first case is realized on the
ultrarelativistic circular orbits of the radius
$r=3m(1+\delta),\quad 0<\delta\ll 1$ ($m$ is the Schwarzschild
mass; in the system of units used here, the velocity of light in
vacuum and the gravitational constant are equal to 1).

The purpose of this paper is to draw attention to a possible case
of the nongeodesic synchrotron radiation caused by the particle's
spin interacting with the gravitational field. That is, when in
addition to the gravitational no other fields or factors are
required. We hope that our investigation will be of interest both
in the context of the recent investigations of the gravitational
radiation of a spinning particle near black holes \cite{3,4,5}, as
well as for general estimates of the nongravitational synchrotron
radiation \cite{6}.

We shall use the Mathisson-Papapetrou equations describing the
motion of a spinning test particle (body) in a gravitational field
\cite{7,8}
\begin{equation}\label{1}
\frac D {ds} \left(Mu^\lambda + u_\mu\frac {DS^{\lambda\mu}}
{ds}\right)= -\frac {1} {2} u^\pi S^{\rho\sigma}
R^{\lambda}_{\pi\rho\sigma},
\end{equation}
\begin{equation}\label{2}
\frac {DS^{\mu\nu}} {ds} + u^\mu u_\sigma \frac {DS^{\nu\sigma}}
{ds} - u^\nu u_\sigma \frac {DS^{\mu\sigma}} {ds} = 0,
\end{equation}
where $u^\lambda$ is the 4-velocity of a spinning particle,
$S^{\mu\nu}$ is the tensor of spin, $M$ and $D/ds$ are,
respectively, the mass and the covariant derivative (here and in
the following, greek indices run 1, 2, 3, 4 and latin indices 1,
2, 3; the signature of the metric $(-,-,-,+)$ is chosen). For the
description of the center of mass of the test particle, equations
(\ref{1}), (\ref{2}) are supplemented by some relationship.
Because the correct definition of the center of mass for a
spinning particle is a subject of discussion, different
relationships (conditions) are used. Mainly the relationship
\cite{7,9}
\begin{equation}\label{3}
S^{\mu\nu} u_\nu = 0
\end{equation}
or \cite{10,11,12}
\begin{equation}\label{4}
S^{\mu\nu} P_\nu = 0
\end{equation}
are considered where
\begin{equation}\label{5}
P^\nu = Mu^\nu + u_\mu \frac{DS^{\nu\mu}}{ds}
\end{equation}
(the corresponding references can be found in \cite{13}).

It is shown that equations (\ref{1}), (\ref{2}) are the classical
limit of the Dirac equation in a gravitational field \cite{14}.

\section {Circular orbits of highly relativistic spinning
particles in a Schwarzschild field}

We begin by considering  equations (\ref{1})-(\ref{3}). By
equation (\ref{3}) the components $S^{i4}$ can be expressed
through $S^{i4}$:
\begin{equation}\label{6}
 S^{i4}=\frac{u_k}{u_4}S^{ki}.
\end{equation}
Using (\ref{6}), the components $S^{i4}$ can be eliminated from
equations (\ref{1}), (\ref{2}). Instead of three independent
components $S^{ik}$ (the tensor $S^{\mu\nu}$ is antisymmetric in
$\mu, \nu$) it is appropriate to use their linear combinations
\begin{equation}\label{7}
 S_i=\frac{1}{2}\sqrt{-g}\varepsilon_{ikl} S^{kl}
\end{equation}
where $g$ is the determinant of the metric tensor,
$\varepsilon_{ikl}$ is the spatial Levi-Civita symbol. The direct
calculation shows that the 3-component value $S_i$ has the
3-vector properties relative to the coordinate transformations
$x^{'i}=x^{'i}(x^1, x^2, x^3),\quad
 x^{'4}=x^4$. The similar spin
3-vector was used in \cite{15, 16} for the concrete metric. In
many papers the 4-vector of spin $s_\lambda$ is considered where
\begin{equation}\label{8}
 s_\lambda=\frac{1}{2}\sqrt{-g}\varepsilon_{\lambda\mu\nu\sigma}u^\mu S^{\nu\sigma}
\end{equation}
($\varepsilon_{\lambda\mu\nu\sigma}$ is the Levi-Civita symbol).
By equations (\ref{6})-(\ref{8}) we have the relationship between
$S_i$ and $s_\lambda$
$$
S_i=u_is_4-u_4s_i.
$$

Using relationships
(\ref{3}), (\ref{7}) three independent equations from (\ref{2})
may be written in the form
\begin{equation}\label{9}
u_4\dot S_i-\dot u_4 S_i+2\left(\dot u_{[4} u_{i]}-u^\pi u_\rho
\Gamma^\rho_{\pi [4} u_{i]}\right)S_k u^k+2S_n\Gamma^n_{\pi[4}
u_{i]} u^\pi =0
\end{equation}
where a dot denotes differentiation with respect to the proper
time $s$, and square brackets denote antisymmetrization of
indices; $\Gamma^\rho_{\pi 4}$ are the Christoffel symbols. (It is
known that among six equations from (2) there are only three
independent equations).

Let us consider (\ref{9}) for the Schwarzschild metric using the
standard coordinates $x^1=r,\quad x^2=\theta,\quad
x^3=\varphi,\quad x^4=t$. It is easy to check that three equations
from (\ref{9}) have a partial solution with $\theta = \pi/2, \quad
S_1\equiv S_r=0, \quad S_3\equiv S_\varphi=0$ and the relationship
for the nonzero component of the spin 3-vector $S_2\equiv
S_\theta$ is
\begin{equation}\label{10}
S_2=ru_4S_0
\end{equation}
where $S_0$ is the constant of integration. The physical meaning
of this constant is the same as in the general integral of the
Mathisson-Papapetrou equations \cite{17}
\begin{equation}\label{11}
S_0^2=\frac12S_{\mu\nu}S^{\mu\nu}.
\end{equation}
We stress that relationship (\ref{10}) is valid for any equatorial
motions ($\theta=\pi/2$) when spin is orthogonal to the motion
plane ($S_k u^k=0$).

The possible equatorial orbits of a spinning particle are
described by equation (\ref{1}). In the following we shall confine
ourselves to the case of the circular orbits.

For the investigation of conditions of existence of equatorial
circular orbits of a spinning particle in a Schwarzschild field we
shall use equations (\ref{1}), (\ref{10}) and the general
relationship for the velocity components $u_\mu u^\nu = 1$. From
the geodesic equations in this field follow the algebraic
relationships determining the dependence of the velocity of a
particle without spin on the radius of the circular orbit.
Similarly, from equation (\ref{1}) we obtain the relationship for
the equatorial circular orbits of a spinning particle in a
Schwarzschild field
$$
u_\perp^3\beta\left(1-\frac{3m}{r}\right)^2-u_\perp^2\left(1-\frac{2m}{r}\right)
\left(1-\frac{3m}{r}\right)
$$
\begin{equation}\label{12}+
u_\perp\beta\frac{m}{r}\left(2-\frac{3m}{r}\right)+\frac{m}{r}\left(1-\frac{2m}{r}\right)=0
\end{equation}
where $u_\perp = r\dot\varphi$ and $\beta\equiv S_2/Mr^2$. In the
case when the spin is equal to 0 (i.e. $\beta = 0$), from equation
(\ref{12}) follows the known second-order equation relatively
$u_\perp$ for the geodesic circular orbits together with the
conclusion that such orbits exist at $r>3m$ only. If $\beta \ne
0$, equation (\ref{12}) can be used for analysis of the possible
circular motions of a spinning particle.

First of all, we see that at $r=3m$ equation (\ref{12}) has the
single solution
\begin{equation}\label{13}
u_\perp=-\frac{1}{3\beta}.
\end{equation}
For the estimate of the absolute value $|u_\perp|$, in (\ref{10})
we take into account the condition for a test spinning particle
\cite{18}
\begin{equation}\label{14}
\frac{|S_0|}{Mm}\equiv \varepsilon\ll 1.
\end{equation}
By equations (\ref{10}), (\ref{13}), (\ref{14}) we have
\begin{equation}\label{15}
|u_\perp|=\frac{3^{1/4}}{\sqrt{\varepsilon}}.
\end{equation}
Hence $u_\perp^2\gg 1$, that is, for the motion on the circular
orbit $r=3m$, the velocity of a spinning particle must be
ultrarelativistic, the more ultrarelativistic the smaller is the
ratio of spin to mass of the particle.

Before the analysis of the solutions of equation (\ref{12}) at
$r\ne 3m$, we stress the following important fact. If instead of
equation (\ref{1}) its shortened variant
\begin{equation}\label{16}
M\frac D {ds} u^\lambda=-\frac {1} {2} u^\pi S^{\rho\sigma}
R^{\lambda}_{\pi\rho\sigma}
\end{equation}
is taken into account [Equation (\ref{16}) is often considered in
different papers], then for the description of the possible
circular orbits of a spinning particle in a Schwarzschild field
follows the equation
\begin{equation}\label{17}
u_\perp^2\left(1-\frac{3m}{r}\right)-u_\perp\frac{3m}{r}\beta
-\frac{m}{r}=0.
\end{equation}
Equation (\ref{17}) at $r=3m$ has the single solution $u_\perp =
-1/3\beta$ which coincides with (\ref{13}). That is, the orbit
$r=3m$ is particular in the sense that the corresponding solution
for $u_\perp$ is common for equations (\ref{12}) and (\ref{17}).

It is easy to see that according to (\ref{12}) the circular
ultrarelativistic orbits of a spinning particle exist in the small
neighborhood of the value $r=3m$ as well, as for $r>3m$ and for
$r<3m,$ in contrast with the geodesic circular orbits. In
particular, if $r=3m(1+\delta),\quad |\delta|\ll \varepsilon$,
then instead of (\ref{13}), (\ref{15}) from (\ref{12}) we obtain
\begin{equation}\label{18}
u_\perp=-\frac{1}{3\beta}\left(1-\frac{3^{3/2}}{2}\frac{\delta}{\varepsilon}\right),
\end{equation}
\begin{equation}\label{19}
|u_\perp|=\frac{3^{1/4}}{\sqrt{\varepsilon}}
\left(1-\frac{3^{3/2}}{2}\frac{\delta}{\varepsilon}\right).
\end{equation}
 Here we stress that for the concrete orbit of
radius $r=3m(1+\delta)$ with $\delta > 0$ the conditions of
realization of the motion for the spinning test particle and the
particle without spin are essentially different. Indeed, according
to the geodesic equations, for the motion on such an orbit the
particle must posses the velocity corresponding to the
relativistic Lorentz factor of order $1/\sqrt{\delta}.$ Whereas,
according to (\ref{19}), for the motion on the orbit of the same
radius at $0<\delta\ll \varepsilon$ the velocity of the spinning
particle must correspond to the Lorentz factor of order
$1/\sqrt{\varepsilon}$. That is, the ratio of the Lorentz factor
values for these cases is equal to $\sqrt{\varepsilon/\delta}$ and
at very small $\delta$ becomes considerably greater than 1.
Formally at $\delta = 0$ this ratio is equal to $\infty$. Actually
it means that a particle without spin and with nonzero mass of any
velocity close to the velocity of light, starting in the
tangential direction from the position $r=3m,$ will fall on the
horizon surface within a finite proper time. On the other hand, a
spinning particle will remain indefinitely on the circular orbit
$r=3m$ due to the interaction of its spin with the gravitational
field. This interaction compensates the usual (geodesic)
attraction. Certainly, this picture holds in the ideal case, when
perturbations are neglected, because the circular orbits in the
neighbor of $r=3m(1+\delta)$ are unstable both for a particle
without spin and for a spinning particle. In reality, one is
dealing with fragments of trajectories close to the corresponding
circular orbits.

We point out that equation (\ref{12}) at $m=0$ has (beside the
trivial case $u_\perp=0$) the solution
\begin{equation}\label{20}
u_\perp=\frac{1}{\beta}.
\end{equation}
Relationship (\ref{20}) describes the known Weyssenhoff circular
orbits \cite{19} of the non-proper center of mass ( by the
terminology of C.M\"oller \cite{20}) of a body rotating in the
Minkowski spacetime. It is important that at the fixed sign of
$S_2$ (i.e. when the direction of the inner rotation is fixed) the
signs of $u_\perp$ in (\ref{18}) and (\ref{20}) are opposite. It
means that the circular orbits in a Schwarzschild field which are
described by equations (\ref{13}), (\ref{15}), (\ref{18}),
(\ref{19}) do not pass into the Weyssenhoff orbits at $m \to 0$,
and the physical nature of these orbits is different. This
conclusion is also supported by the fact that the Weyssenhoff
orbits correspond to equation (\ref{12}) and do not correspond to
equation (\ref{17}), whereas solution (\ref{13}) is common for
equations (\ref{12}) and (\ref{17}). In this context we emphasize
the following fact. Though equations (\ref{13}), (\ref{18})
describe the solutions of equations (\ref{1}), (\ref{2}) under
condition (\ref{3}), these solutions satisfy (in the main
appoximation) condition (\ref{4}) as well. Indeed, from (\ref{3}),
\ref{6}), (\ref{7}) we have the nonzero components of $S^{\mu\nu}$
for a spinning particle on the circular orbits:
\begin{equation}\label{21}
S^{13}=-S^{31}=\frac{S_2}{r^2}, \qquad
S^{14}=-S^{41}=\frac{u_3S_2}{u_4 r^2}.
\end{equation}
Taking into account expressions (\ref{5}), (\ref{10}), (\ref{13}),
(\ref{14}) and (\ref{21}) we obtain
$$
P^1=0, \qquad P^2=0,
$$
$$
P^3=-\frac{3^{-3/4}M}{m\sqrt{\varepsilon}}\left(1-\frac{11\sqrt{3}}{18}\varepsilon\right)signS_0,
$$
\begin{equation}\label{22}
P^4=\frac{3^{3/4}M}{\sqrt{\varepsilon}}\left(1+\frac{7}{6\sqrt{3}}\varepsilon\right).
\end{equation}
By equations (\ref{21}), (\ref{22}) it is easy to check that in
this case
$$
S^{2\nu}P_\nu=0, \quad S^{3\nu}P_\nu=0, \quad S^{4\nu}P_\nu=0,
$$
\begin{equation}\label{23}
 S^{1\nu} P_\nu=\frac{1}{9}\varepsilon^2 mM^2.
\end{equation}
So, the expression $S^{1\nu} P_\nu$, though is not equal to {0},
is proportional to $\varepsilon^2$. We also stress that in the
practical situations $mM^2\ll 1$ (for example, in the system used
$m$ for the Sun is equal to $5\cdot 10^{-6}$, and $M$ for an
electron is equal to $2\cdot 10^{-66}$). That is, the value of
$S^{1\nu} P_\nu$ is much less than 1. In this sense the above
considered solutions of equations (\ref{1}), (\ref{2}) determined
by (\ref{13}), (\ref{18}) are practically common for conditions
(\ref{3}) and (\ref{4}). It is known that under condition
(\ref{4}) equations (\ref{1}), (\ref{2}) do not have solutions of
the Weyssenhoff type.

It is interesting that equation (\ref{12}) has a single real
solution for any fixed $r$ from the interval $2m<r<3m$ which is
not located in the small neighbor of $r=3m$. The corresponding
dependence on $r$ is
\begin{equation}\label{24}
u_\perp=\frac{1}{\beta}\left(1-\frac{2m}{r}\right)\left(1-\frac{3m}{r}\right)^{-1}
\left[1-\frac{3m}{r}\left(1-\frac{2m}{r}\right)^{-1}\beta^2\right]
\end{equation}
where, as above, we take into account the condition $\beta^2\ll
1$. That is, as in cases (\ref{13}), (\ref{18}), the value
$u_\perp$ is ultrarelativistic. It is important that (because
$1-3m/r < 0$) the sign of $u_\perp$ is opposite to the sign of
$\beta$, as in cases (\ref{13}), (\ref{18}) and in contrast with
the Wyessenhoff case (\ref{20}). By this feature the circular
orbits from the interval $2m<r<3m$, described by expression
(\ref{24}), can be interpreted as such whose nature is the same as
for orbits (\ref{13}), (\ref{18}). Namely, these orbits are caused
by the interaction of the spin with the gravitational field.
According to the analysis from \cite{21}, for any (circular or
not) motion of a spinning particle in a Schwarzschild field, this
interaction has specific spin-orbit properties. At the same time,
it is easy to see that equation (\ref{12}) at $r>3m$ has the
solution for $u_\perp$ which at $m/r \to 0$ passes into
(\ref{20}), i.e. this solution is of the Weyssenhoff type.

The behaviour of a spinning test particle on circular orbits in
the Schwarzschild and Kerr fields (mainly the clock effect) was
studied in latest papers \cite{22} at different conditions for the
Mathisson-Papapetrou equations. Different types of orbits of the
spinning particles in a Schwarzschild field were investigated in
\cite{23} on the base of the Mathisson-Papapetrou equations at
condition (\ref{4}).

\section {Numerical estimates and conclusions}

So, according to the Mathisson-Papapetrou equations a spinning
test particle can moves on ultrarelativistic equatorial circular
orbits with the values of the radial coordinate $r$ from the
region near $r=3m$ in a Schwarzschild field. The spin of a
particle on these orbits is orthogonal to the motion plane, and,
by (\ref{15}), (\ref{19}), the particle's velocity is more higher,
the smaller is the ratio of spin to mass of the particle.

The existence of ultrarelativistic circular orbits of a spinning
particle in a Schwarzschild field, which differ from geodesic
circular orbits, probably, can be discovered in the synchrotron
radiation of protons or electrons. Let us estimate the value
$\varepsilon=|S_0|/Mm$ for these particles in the cases when the
Schwarzschild source is 1) a black hole of mass that is equal to
three of the Sun's mass, and 2) a massive black hole of mass that
is equal to $10^8$ of the Sun's mass. In the first case, taking
into account the numerical values of $S_0, M, m$ in the system
used, we have for protons and electrons correspondingly
$\varepsilon_p \approx 2\cdot 10^{-20},\quad \varepsilon_e \approx
4\cdot 10^{-17}$, whereas in the second case $\varepsilon_p
\approx 7\cdot 10^{-28},\quad \varepsilon_e \approx 10^{-24}$.
Because for the motion on above considered circular orbits the
spinning particles must possess the velocity corresponding
(according to equations (\ref{15}), (\ref{19})) to relativistic
Lorentz $\gamma$-factor of order $1/\sqrt{\varepsilon}$, in the
first case we obtain $\gamma_p \approx 7\cdot 10^9,\quad \gamma_e
\approx 2\cdot 10^8$, and in the second case $\gamma_p \approx
4\cdot10^{13}$,\quad $\gamma_e \approx 10^{12}$.

As we see, in the case of a massive black hole the necessary
values of $\gamma_p$ and $\gamma_e$ are too high even for
extremely relativistic particles from the cosmic rays. Whereas if
the Schwarzschild source is a black hole of mass that is equal to
3 of the Sun's mass, some particles may move on the circular
orbits considered above. Analysis of a concrete model, closer to
the reality, remains to be carried out. Here we point out that by
the known general relationships for the electromagnetic
synchrotron radiation in the case of protons or electrons on the
considered circular orbits we obtain the values from the gamma-ray
range. By the known features of the synchrotron radiation and the
null geodesic trajectories, the part of the synchrotron radiation
which arises from the orbits with $r<3m$ cannot leave the surface
of radius $r=3m$ and will fall on the black hole.

It is known that he highly relativistic circular orbits of a
spinless test particle are of importance in the classification of
all possible geodesic orbits in a Schwarzschild field. The results
of the present work show the significance of the ultrarelativistic
circular orbits of a spinning particle for the description of all
possible essentially nongeodesic trajectories following from the
Mathisson-Papapetrou equations. First of all, namely on the
circular or close to the circular ultrarelativistic orbits the
spinning particle feels the maximal effect of the gravitational
spin-orbit interaction. It would therefore be interesting to study
the dynamics of the spinning particle deviating from the above
considered circular orbits. This is our purpose for the next
paper. Naturally, the problem of considerable importance is the
search of the physical situations where the phenomena connected
with different types of the ultrarelativistic motions of the
spinning particles can be discovered.


\vskip 3mm

\newpage
\section*{References}


\begin{thebibliography}{10}
\bibitem{1} Breuer R A, Chrzanowski P L, Hughes H G and Misner C W
1973 \textit{Phys. Rev.} D \textbf{8} 4309
\bibitem{2} Chrzanowski P
L and Misner C W 1974 \textit{Phys. Rev.} D \textbf{10} 1701
\bibitem{3} Saijo M, Maeda K, Shibata M and Mino Y 1998 \textit{Phys.
Rev.} D \textbf{58} 064005
\bibitem{4} Suzuki S and Maeda K 2000
\textit{Phys. Rev.} D \textbf{61} 024005
\bibitem{5} Mohseni M and
Sepangi H 2000 \textit{Class. Quantum Grav.} \textbf{17} 4615
\bibitem{6} Ellis J, Mavromatos N E, Nanopoulos D V and Sakharov A S
2003 \textit{Preprint} astro-ph/0309144
\bibitem{7} Mathisson M 1937
\textit{Acta Phys. Pol.} \textbf{6} 163
\bibitem{8} Papapetrou A 1937
\textit{Proc. R. Soc.} A \textbf{209} 248
\bibitem{9} Pirani F A E
1956 \textit{Acta Phys. Pol.} \textbf{15} 389
\bibitem{10} Tulczyjew W
1959 \textit{Acta Phys. Pol.} \textbf{18} 393
\bibitem{11} Dixon W
1970 \textit{Proc. R. Soc.} A \textbf{314} 499
\bibitem{12} Tod K P and de Felice F
1976 \textit{IL Nuovo Cimento} \textbf{34} 365
\bibitem{13} Semerak O
1999 \textit{Mon. Not. R. Astron. Soc.}  \textbf{308} 863
\bibitem{14}
Wong S 1972 \textit{Int. J. Theor. Phys.}  \textbf{5} 221

Kannenberg L 1977 \textit{Ann. Phys.(N.Y.)} \textbf{103} 64

Catenacci R and Martellini M 1977 \textit{Lett. Nuovo Cimento}
\textbf{20} 282

Audretsch J 1981 \textit{J. Phys.} A  \textbf{14} 411

Gorbatsievich A 1986 \textit{Acta Phys. Pol.} B  \textbf{17} 111

Barut A and Pavsic M 1987 \textit{Class. Quantum Grav.} \textbf{4}
41
\bibitem{15} Micoulaut R 1967 \textit{Z. Phys.} \textbf{206} 394
\bibitem{16} Carmeli M, Charach Ch and Kaye M 1977 \textit{Phys. Rev.} D \textbf{15} 1501
\bibitem{17} Mashhoon B 1971 \textit{J. Math. Phys.} \textbf{12}
1075
\bibitem{18} Wald R 1972 \textit{Phys. Rev.} D \textbf{6} 406
\bibitem{19} Weyssenhoff J and Raabe A 1947 \textit{Acta. Phys. Pol.}
\textbf{9} 7
\bibitem{20} M\"oller C 1949 \textit{Commun. Dublin Inst.
Advan. Studies} A  \textbf{5} 3

M\"oller C 1949 \textit{Ann. Inst. Henri Poincare} \textbf{11} 251

\bibitem{21}  Plyatsko R, Bilaniuk O 2001 \textit{Class. Quantum Grav.}
\textbf{18} 5187

\bibitem{22}  Bini D, de Felice F and Geralico A  2004
\textit{Preprint} gr-qc/0410082

Bini D, de Felice F and Geralico A
2004 \textit{Preprint} gr-qc/0410083
\bibitem{23} Suzuki S and Maeda K 1997
\textit{Phys. Rev.} D \textbf{55} 4848

\end{thebibliography}
\end{document}